\documentclass[twocolumn,preprintnumbers,superscriptaddress,endnote,nofootinbib,prd]{revtex4}

\usepackage{graphicx}

\usepackage[hypertex]{hyperref}
\newcommand{\vev}[1]{\langle {#1} \rangle}
\newcommand{\lsim}{\lesssim}
\newcommand{\gsim}{\gtrsim}

\begin{document}

\pagestyle{plain}

\preprint{MADPH-04-1397}

\title{Models of Neutrino Mass with a Low Cutoff Scale}

\author{Hooman Davoudiasl}
\affiliation{Department of Physics, University of Wisconsin,
  Madison, WI 53706}
\affiliation{School of Natural Sciences, Institute for Advanced Study,
  Princeton, NJ 08540}

\author{Ryuichiro Kitano}
\affiliation{School of Natural Sciences, Institute for Advanced Study,
  Princeton, NJ 08540}

\author{Graham D. Kribs}
\affiliation{School of Natural Sciences, Institute for Advanced Study,
  Princeton, NJ 08540}
\affiliation{Department of Physics, University of Oregon,
  Eugene, OR 97403\footnote{On leave of absence.}}

\author{Hitoshi Murayama}
\affiliation{Department of Physics, University of California,
  Berkeley, CA 94720}
\affiliation{Theoretical Physics Group, Lawrence Berkeley National Laboratory,
  Berkeley, CA 94720}
\affiliation{School of Natural Sciences, Institute for Advanced Study,
  Princeton, NJ 08540}


\begin{abstract}
  In theories with a low quantum gravity scale, global symmetries
  are expected to be violated, inducing excessive proton decay or
  large Majorana neutrino masses.  The simplest cure is to impose
  discrete gauge symmetries, which in turn make neutrinos massless.
  We construct models that employ these gauge symmetries while
  naturally generating small neutrino masses.  Majorana (Dirac)
  neutrino masses are generated through the breaking of a discrete
  (continuous) gauge symmetry at low energies, {\it e.g.}\/,
  2~keV--1~GeV\@.  The Majorana case predicts $\Delta N_\nu \simeq 1$
  at BBN, neutrinoless double beta decay with scalar emission, 
  and modifications to the CMB anisotropies from domain walls 
  in the universe as well as providing a possible Dark Energy candidate.  
  For the Dirac case, despite the presence of a new light gauge boson, all
  laboratory, astrophysical, and cosmological constraints can be
  avoided.

\end{abstract}
\maketitle

\section{Introduction}

The hierarchy between the electroweak scale and the Planck scale is one
of the most puzzling aspects of the Standard Model (SM).  On the one
hand, allowing the cutoff scale of the SM to be the Planck scale allows
us to naturally understand both the stability of the proton and the
smallness of the neutrino masses through the seesaw mechanism. On the
other hand, an enormous fine-tuning is required to prevent the Higgs
mass from being destabilized by radiative corrections proportional to
the cutoff scale.  Our options appear to be to either live with 
fine-tuning (for example \cite{Davoudiasl:2004be}) or find a dynamical
solution that may well affect the successes of the Standard Model with 
a high cutoff scale.

The most dramatic solution to the hierarchy problem is to lower the
cutoff scale of the SM close to the weak scale.  In scenarios with
large or warped extra dimensions \cite{Arkani-Hamed:1998rs,Randall:1999ee}, 
the cutoff scale of the SM is the fundamental scale
of quantum gravity.  All global symmetries are expected to be violated
by quantum gravity effects, in particular baryon and lepton number are
violated leading to extremely rapid proton decay and way-too-large
Majorana neutrino masses.  There are intrinsically extra dimensional
solutions to these problems
\cite{Arkani-Hamed:1999dc,Grossman:1999ra}, 
for example, by separating fermions
within a fat brane. The solution to the too-large Majorana neutrino
mass problem, however, requires more structure beyond fermion
geography.\footnote{Consider, {\it e.g.}\/, the interesting proposal
  of \cite{Arkani-Hamed:1999dc}, in which U(1)$_{B-L}$ is gauged in
  the bulk.  The needed geography is to separate left-handed neutrinos
  from right-handed neutrinos from U(1)$_{B-L}$ breaking fields in the
  extra dimensional space. It also requires additional scalar fields
  with suitable profiles to stabilize the positions of the fermions.
  Furthermore, the smallest cutoff scale must be parametrically larger
  than the inverse distance separating left-handed and right-handed leptons 
  to ensure the wave-function overlaps are sufficiently small.}

In this paper we consider simple, four-dimensional solutions to the
problems of baryon and lepton number violation. Our proposals are
applicable to models of large extra dimensions, as well as any other
model that lowers the cutoff scale of the SM, violating the global
symmetries. We propose to avoid those problems by imposing new gauge
symmetries in the effective theory that forbid the dangerous
operators. Baryon number and lepton number violation are eliminated (up
to $\Delta B = 3$ and $\Delta L = 3$) by imposing a gauged $Z_9$ baryon
number and a gauged $Z_3$ lepton number.  Similar gauged discrete
symmetries were employed in \cite{Ibanez:1991hv,Babu:2003qh}.

Gauging discrete baryon number eliminates the proton decay
problem. Gauging discrete lepton number causes the neutrinos of
the SM to be massless.  Since the cutoff scale is low, we lose the
usual seesaw mechanism as an explanation for the smallness of
neutrinos masses.  Here we will consider two models to generate
neutrino masses of the experimentally observed size. The first is
to break $Z_3$ lepton number by a small amount, generating small
Majorana neutrino masses. The second is to introduce right-handed
neutrinos and a new U(1) gauge symmetry that is spontaneously
broken at a scale deep in the infrared, generating small Dirac
neutrino masses. We subject our models to severe experimental,
astrophysical, and cosmological constraints and demarcate the
regions of parameter space that survive.  Remarkably, despite the
presence of a new scalars and/or a gauge boson with mass well
below the electroweak scale, we find these models provide viable
solutions to the problems of baryon and lepton number violation
from low cutoff scale operators.

The models we present are similar to those in
Refs.~\cite{Chacko:2003dt,Chacko:2004cz}, in that
neutrino mass originates from a symmetry breaking at low scales,
and hence a phase transition at late times.  There is, however, an
important difference.  While they have focused on nearly massless
pseudo-Nambu-Goldstone bosons due to spontaneously broken global
symmetries, we insist on all symmetries to be gauged and we do not
have any nearly massless bosons.  Therefore, in general, the
phenomenological constraints are correspondingly quite different.
We note that models with a low cutoff scale based on discrete
symmetries, but without dynamics in the infrared regime we consider,
have been studied in Ref.~\cite{Babu:2003qh}.

The paper is organized as follows.  In the next section, we introduce
flavor-blind discrete gauge symmetries that ensure the absence of
too-rapid proton decay and Majorana neutrino masses.  In section
\ref{sec:Majorana}, we break the discrete gauge symmetry spontaneously
and induce the Majorana neutrino masses at the required size.  We show
that the constraints from the Big-Bang Nucleosynthesis (BBN) as well as the
neutrinoless double beta decay with an additional scalar emission are
marginal.  The network of domain walls is an inevitable prediction,
which naturally satisfies the constraint from cosmic microwave
anisotropy and can be a candidate for Dark Energy.  We discuss the
Dirac case in section \ref{sec:Dirac}.  We first show that an
additional discrete gauge symmetry can suppress the Dirac neutrino
mass, but the predicted domain wall network requires a high cutoff
scale.  We then focus on a new continuous gauge symmetry, which
necessarily predicts a new very light gauge boson, and show
that all laboratory, astrophysical, and cosmological constraints can
be satisfied.  Finally we comment on the hierarchy between the
neutrino mass scale and the fundamental scale in section
\ref{sec:hierarchy}, and conclude in section \ref{sec:conclusion}.

\section{Protecting Matter\label{sec:protecting}}

The most pressing problems of lowering the cutoff scale in the SM
are fast proton decay and too large Majorana neutrino masses. We
propose a set of two discrete symmetries; the SM quarks transform
under one symmetry, and leptons transform under the other. Here,
we have assumed that the chiral fermion content is the same as in
the minimal SM\@.  The discrete symmetries are assumed to be
flavor-blind (although we comment on a non-flavor-blind variation
that suppresses $\mu \to e\gamma$ in Appendix \ref{alternative-appendix}.). 
To ensure that the gauged
discrete symmetries are not badly broken at the electroweak scale,
the Higgs is taken to be neutral. Allowing ordinary Yukawa
couplings enforces the conditions $Z_A^q(Q) = - Z_A^q(u^c) = -
Z_A^q(d^c)$ and $Z_B^\ell(L) = - Z_B^\ell(e^c)$. This leaves us
with
\begin{center}
\begin{tabular}{l|ccccc}
            &  $Q$  &  $u^c$  &  $d^c$  &  $L$  &  $e^c$ \\ \hline
$Z_A^q$     &  $a$  &   $-a$  &   $-a$  &  $0$  &    $0$ \\
$Z_B^\ell$  &  $0$  &    $0$  &    $0$  &  $b$  &   $-b$
\end{tabular}
\end{center}
The $Z_A^q$-$SU(3)$-$SU(3)$ and $Z_A^q$-grav-grav anomalies are
canceled for any $a,A$. The $Z_A^q$-$SU(2)$-$SU(2)$ anomaly
requires $9 a = 0 \; {\rm mod} \; A$. We choose $(A,a) = (9,1)$.
Similarly the $Z_B^\ell$-$SU(2)$-$SU(2)$ and $Z_B^\ell$-grav-grav
anomalies require $3 b = 0 \; {\rm mod} \; B$, and hence $(B,b) =
(3,1)$. Mixed $U(1)$-discrete anomalies can be satisfied by redefining
the $U(1)$ normalization, and pure discrete anomalies can be satisfied
by adding additional massive matter with suitable 
(fractional) discrete charges \cite{Ibanez:1991hv, Banks:1991xj}.
Note that these discrete symmetries, up to certain charge redefinitions,
have been proposed before in supersymmetric \cite{Ibanez:1991hv}
and non-supersymmetric \cite{Babu:2003qh} contexts.  In particular,
our $Z_3^\ell$ is equivalent to $L_3$ of \cite{Ibanez:1991hv}.  Our
$Z_9^q$ is equivalent to $R_3^{-3} L_3^{-3} Y$ of \cite{Ibanez:1991hv}
where $R_3^{-3},L_3^{-3}$ is understood to mean $R_3,L_3$ promoted to
$R_9,L_9$ with all fields' discrete charge multiplied by $-3$, and
$Y$ refers to the $Z_9$ subgroup of $U(1)_Y$.

The lowest dimension operators leading to baryon and lepton number
violation are schematically
\begin{equation}
\frac{Q^9 L^3}{\Lambda^{14}} \quad + \; \rm{others...}
\end{equation}
The most conservative assumption is to require this $\Delta B=3$ process
leads to a multi-nucleon lifetime that is longer than the proton lifetime,
\begin{equation}
\tau = \frac{\Lambda^{28}}{m_{nuc}^{29}} > 10^{33} \; \rm{yrs}
\end{equation}
and thus $\Lambda \gsim 200 m_{nuc}$.  This constraint is easily
satisfied for the cutoff scales we consider in this paper.
Similarly Majorana neutrino mass is completely forbidden so
long as $Z_3^\ell$ is exact.

\section{$Z_3$ Breaking and Majorana Neutrinos\label{sec:Majorana}}

With exact $Z_3^\ell$ lepton number symmetry, Majorana neutrino masses
arising from cutoff scale operators are forbidden. To generate non-zero
neutrino masses, of Majorana type, the $Z_3$ symmetry must be broken.
We show that a low breaking scale is needed, of order 2--400~keV, to
give a consistent framework.  Discrete symmetry breaking leads to domain
walls, that can induce a too large anisotropy in the CMB\@.  However, a
frustrated network of domain walls may avoid this constraint and even
provide a candidate for dark energy.

To generate neutrino masses that are purely Majorana, $Z_3^\ell$ is
broken by a complex scalar field $\chi$ with $Z_3^\ell$ charge that
acquires a vev.  Without loss of generality we can choose the $Z_3^\ell$
charge of $\chi$ to be equal to the leptons.  This allows us to write
the dimension-6 operator
\begin{equation}
c \frac{\chi L H L H}{\Lambda^2} \;
\label{Majorana-op}
\end{equation}
with a coefficient that we assume takes on a natural value, $c \sim
1$. The other nontrivial choice of discrete charge for $\chi$, namely
$Z_3^\ell(\chi)=2$, leads to the same dimension-6 operator upon simply
replacing $\chi \to \chi^*$ in the above. The effective Yukawa
coupling is
\begin{equation}
  \label{eq:9}
  g \nu \nu \chi, \qquad g = c \frac{v^2}{\Lambda^2} ,
\end{equation}
giving a neutrino mass
\begin{equation}
m_\nu = g \langle \chi \rangle = (0.06 \; {\rm eV}) \, c \, \left(
\frac{\langle \chi \rangle}{2 \; {\rm keV}} \right) \, \left(
\frac{30 \; {\rm TeV}}{\Lambda} \right)^2 \; . \label{numass-eq}
\end{equation}

An important constraint on the Yukawa coupling $g$ arises from
$0\nu\beta\beta\chi$ process where the $\chi$ is emitted in the
$0\nu\beta\beta$ transition \cite{Georgi:1981pg}.
Assuming a natural potential where $m_{\chi} \sim \langle \chi
\rangle$, a physical $\chi$ is easily emitted in the nuclear
transition $0\nu\beta\beta\chi$ with no kinematic suppression.
Experimental constraints on this
process have determined that the effective coupling of $\chi \nu
\nu$ must be smaller than $3 \times 10^{-5}$
\cite{Bernatowicz:1992ma}.  This implies $\Lambda > 30$~TeV,
taking $c=1$, independent of any other assumptions.
The smallest neutrino mass that can explain the
atmospheric neutrino oscillation data is approximately
$\sqrt{\Delta m_{\rm atm}^2} \simeq 0.06$ eV, which implies
$\vev{\chi} > 2$ keV from Eq.~(\ref{numass-eq}).

Spontaneous breaking of a discrete symmetry can lead to a domain
wall problem.  A network of domain walls typically ``scales,''
\emph{i.e.}, the walls stretch and simplify their configuration
and there is basically always only one (or a few) of the walls
within the horizon. The energy density of the walls is
approximately $\sigma/L$, where $\sigma \sim \vev{\chi}^3$ is the
wall tension and $L \sim H^{-1}$ is the typical inter-wall
distance.  This implies the energy density of the walls is
\begin{equation}
\Omega_{\rm wall} \sim \frac{\sigma H}{M_{Pl}^2 H^2} 
\sim 10^{-12} \left( \frac{\vev{\chi}}{2 \; {\rm keV}} \right)^3
\end{equation}
which implies the upper bound $\vev{\chi} < 25$ MeV such that
the walls do not overclose the universe.
However, there is a more stringent constraint from the cosmic
microwave background (CMB) photons being affected by the walls
themselves.  The simple estimate is that $\Delta T/T \sim G_N \sigma L$
\cite{Friedland:2002qs}.  The observational bound $\Delta T/T
\lesssim 10^{-5}$ requires $\sigma \sim \langle \chi
\rangle^3\lesssim (400~{\rm keV})^3$, which gives the upper bound
$\Lambda \lesssim 400$~TeV\@.  Smaller breaking scales may still leave 
an interesting imprint on the power spectrum or galaxy distribution
functions.  Unfortunately we are not aware of a detailed study of
the imprint of a domain wall network due to a late time phase
transition, and we cannot make any further quantitative
statements.  

It is possible that the domain wall network is {\it frustrated}\/,
namely that the network is so complicated that it cannot simplify
its configuration easily and the configuration expands without
simplifying.  It may happen if $Z_3$ is embedded into a larger
non-Abelian discrete group.  In this case, the domain wall network
can behave as dark energy with $w=-2/3$.  Even though the current
data do not prefer this possibility, analyses without priors
actually allow it \cite{Conversi:2004pi}, and it will surely be
interesting to investigate this possibility in forthcoming
supernovae studies such as SNAP\@.

There is a prediction on the effective number of neutrinos at the
time of BBN\@. The $\chi$ particle can be
generated in early universe through processes such as $\nu \nu \to
\chi$, $\nu \nu \to \chi \chi$, and $\nu \nu \to \chi \chi \chi$.
Although the $\nu \nu \to \chi \chi \chi$ process has the largest
interaction rate for reasonable range of $\Lambda$, we can give a
parameter independent conclusion with the former two processes.
The interaction rate of those processes at temperature $T$ is
roughly given by
\begin{eqnarray}
  \Gamma \sim \frac{m_\nu^2}{16 \pi T}\ .
\end{eqnarray}
Here, we have used the natural assumption that $m \sim m_\chi \sim
\langle \chi \rangle$, with $m$ being the coefficient of the
three-point coupling term $m \chi^3$.
Since the interaction rate $\Gamma$ at $T = T_{BBN} \sim 1$~MeV is
larger than the expansion rate $H \sim T_{BBN}^2/ M_{Pl}$ for $m_\nu
\gtrsim 10^{-4}$~eV, the production process is in thermal equilibrium
and thus $\Delta N_\nu$, the effective number of neutrinos minus three,
is predicted\footnote{There are logical possibilities to
avoid such large deviation. For example, small cubic and quartic
couplings $\lambda$ of $\chi$ such as $m \lesssim 10^{-3} \langle \chi
\rangle$ and $\lambda \lesssim 10^{-6}$ may suppress the interaction
rate sufficiently.} to be $8/7$.
This is rather large and in apparent disagreement with the bound 
$N_\nu^{\rm eff} < 3.4$ at 95\% CL found 
in Ref.~\cite{Pierce:2003uh}. However,
additional systematic errors in the helium abundance measurements
may allow such a large effective number of neutrinos, albeit the
errors on the PDG central value \cite{Eidelman:2004wy} would have
to be significantly increased, as discussed in
\cite{Pierce:2003uh}.  Alternatively, a large chemical potential for 
$\nu_e$ allows up to $\Delta N_\nu \lesssim 4.1$~\cite{Barger:2003rt}.

One may worry about the erasure of the baryon number at high temperature
due to lepton number violating interactions.  It is indeed important for
relatively large $m ( \sim \langle \chi \rangle )$ because the lepton number
erasure process becomes effective before turning off the sphaleron process.
When we assign lepton number $-2$ for $\chi$, the process which violates
lepton number is $\chi \chi \to \chi \chi^* \chi^*$ through $m\chi^3$
term. This process enters thermal equilibrium when temperature drops
down to
\begin{eqnarray}
 T_* \sim 300~{\rm GeV}
\left(
\frac{m}{300~{\rm keV}}
\right)^{2/3}\ .
\end{eqnarray}
By comparing the critical temperature of the electroweak phase
transition $T_c \sim 300$~GeV, we find it is safe for $m \lesssim
300$~keV\@.

Here, we note that the neutrinos interact among themselves by
exchanging $\chi$ with a cross section $\sigma_{\nu\nu} \sim g^4
T^2/m_\chi^4$.  At the time of decoupling, corresponding to $T
\sim 1$~eV, the neutrino mean free path is longer than the horizon
scale, and therefore neutrinos free-stream.  However, the presence
of extra degrees of freedom below the scale of BBN generates a
shift $\Delta l_k$ in the position of the $k^{th}$ acoustic peak
of the CMB\@.  Using the large $k$ results discussed in
Ref.~\cite{Chacko:2003dt}, for the model discussed in this section,
we get $\Delta l_k = -29.3$; the SM prediction for this quantity
is $-23.3$. Given the sensitivity of the Planck experiment, this
deviation will be measurable in the future.

The $Z_3^\ell$ symmetry could be generalized to a $Z_{3N}$,
with $N$ an integer, where the leptons are assigned charge
$N$ (leaving $Z_{3N}$ non-anomalous as before) while $\chi$
is assigned a unit charge.  In this case neutrino masses come
from even higher dimensional operators such as
\begin{eqnarray}
 \left( \frac{\chi}{\Lambda} \right)^N
\frac{LHLH}{\Lambda} \; .
\end{eqnarray}
The symmetry breaking scale in this case is estimated to be
\begin{eqnarray}
 \langle \chi \rangle
\sim 10^{\frac{1}{2}(9-\frac{19}{N})}~{\rm GeV} \left(
\frac{\Lambda}{30~{\rm TeV}} \right)^{1+\frac{1}{N}}\ ,
\end{eqnarray}
where the neutrino mass was fixed to be $0.1$~eV\@.  By using
$\Lambda \gtrsim 30$~TeV and $\langle \chi \rangle \lesssim
400$~keV from the constraint from the domain wall formation, we
find the only viable possibility is $N=1$.  Hence, our choice of
Eq.~(\ref{Majorana-op}) is unique.

\section{New Gauge Theories and Dirac Neutrinos\label{sec:Dirac}}

In this section, we discuss how to generate small Dirac neutrino
masses.  Obviously we need to introduce right-handed neutrinos.  The
smallness of Dirac neutrino masses can be ensured by either an
additional discrete or a new continuous gauge symmetry.  The former
case, however, has severe cosmological constraints.  The latter case
predicts a new very light gauge boson, yet can satisfy all laboratory,
astrophysical, and cosmological constraints depending on the dimension
of the operator responsible for the neutrino masses.

\subsection{Discrete Gauge Symmetry}
\label{Dirac-discrete}

The unwanted large Dirac Yukawa interaction $H L \nu_R$ can be forbidden
by introducing a new gauge symmetry. We discuss new discrete symmetries
under which only the right-handed neutrino and a scalar field $H_R$
transform. The Dirac neutrino masses are generated after the symmetry
breaking through a higher dimensional operator as in the Majorana
case. We show that the constraints from BBN and the domain wall
formation exclude almost all the range of the cutoff scale $\Lambda$
except for a high cutoff region $\Lambda \sim 10^8$~GeV\@.

Since the right-handed neutrinos $\nu_R$ are gauge singlets, it is easy
to construct a model with satisfying the anomaly-free condition. For
example, a model with $Z_3$ symmetry and the following charge assignment
trivially satisfies the condition.
\begin{center}
\begin{tabular}{r|rrr}
           & $\nu^1_R$ & $\nu^2_R$ & $\nu^3_R$  \\ \hline
$Z_{3}^R$      &   $1$    &   $1$    &   $1$        \\
$Z_3^\ell$ &   $-1$    &   $-1$    &   $-1$       
\end{tabular}
\end{center}
The new $Z_{3}^R$ symmetry forbids Dirac Yukawa
interaction terms. The neutrino masses are generated after the
spontaneous $Z_{3}^R$ breaking by the vev of a Higgs field $H_R$ through
the operator:
\begin{eqnarray}
 {\cal L}^R = \lambda^{\alpha i}
\frac{H_R (H L^\alpha) \nu_R^i}{\Lambda}\ ,
\end{eqnarray}
where we assigned the charges of $H_R$ to be $Z_3^R: -1$, and
$\lambda^{\alpha i}$ are $O(1)$ coefficients.  As we did in the Majorana
case, we can generalize the symmetry to $Z_{3n}^R$ and in this case the
neutrino masses are generated through the interaction terms:
\begin{eqnarray}
 {\cal L}^R = \lambda^{\alpha i}
\left( \frac{H_R}{\Lambda} \right)^n
(H L^\alpha) \nu_R^i \ .
\label{eq:Dirac-op}
\end{eqnarray}
Since this is the general form of the interaction term for any choices
of the discrete symmetry group, we do not specify the gauge group
hereafter and discuss the constraint for each $n$.
The symmetry breaking scale is estimated by fixing the neutrino masses
as follows:
\begin{eqnarray}
 \vev{H_R} \sim 10^{3-\frac{12}{n}} {\rm GeV}
\left( \frac{\Lambda}{1~{\rm TeV}} \right) \ .
\label{eq:HRscale}
\end{eqnarray}

Now we consider the constraint on the cutoff scale $\Lambda$ from 
BBN\@.  The agreement of the light element abundances with the predictions
of BBN theory does not allow more than one extra neutrino species during BBN\@.
In our model, $\nu_R$ are light degrees of freedom during BBN, and hence
should satisfy the constraint $\Delta N_\nu \lesssim 1$.
At the time of the QCD phase transition, any preexisting amount of
$\nu_R$ is diluted by a factor of order 10 and hence naturally satisfies 
the limit if $\nu_R$'s are not in the thermal bath.
Instead, the more important constraint arises from the repopulation of 
$R$-sector particles $\nu_R$ and $H_R$ through the interaction
in Eq.~(\ref{eq:Dirac-op}) after the QCD phase transition.

The simplest constraint arises from dimension six operators such as
\begin{equation}
\frac{(\bar{l} \gamma^\mu l) (\bar{\nu}_R \gamma_\mu \nu_R)}{\Lambda^2}
\label{D6-eq}
\end{equation}
that lead to $\nu_R$ production.  Requiring that this process is 
frozen out at $T \sim 200$~MeV, one finds the constraint 
$\Lambda > 5$~TeV independent of $n$.

$R$-sector particles can also be produced through the $n$-dependent 
operator, Eq.~(\ref{eq:Dirac-op}).  We first consider the case $n=1$.
After electroweak symmetry breaking, neutrinos have a
Yukawa interaction with $H_R$:
\begin{eqnarray}
 {\cal L}_Y = \frac{\lambda^{\alpha i} v}{\Lambda} 
\nu_L^\alpha \nu_R^i  H_R\ .
\label{eq:Yukawa-Dirac}
\end{eqnarray}
This interaction is important for the generation of $\nu_R$ and $H_R$
through the processes $\nu_L \bar{\nu}_L \to \nu_R \bar{\nu}_R$ and
$\nu_L \bar{\nu}_L \to H_R H_R^*$ via $t$-channel $H_R$ and $\nu_R$
exchange diagrams.
The abundance of these $R$-sector particles is approximately
given by
\begin{eqnarray}
{\xi_R} \equiv  \frac{n_R}{T^3} 
\sim 
{\rm min}\left( \frac{v^4}{4 \pi \Lambda^4} \frac{M_P}{T}
, 1 \right)\ ,
\end{eqnarray}
where $n_R$ is the number density of the $R$-sector particles.
Even if there is no preexisting abundance of $R$-sector particles, 
the small amount produced by the above processes triggers more efficient
production processes such as $\nu_L \nu_R \to H_R H_R^* H_R^*$ and
$\nu_L H_R \to \bar{\nu}_R H_R H_R^*$ through the Yukawa coupling in
Eq.~(\ref{eq:Yukawa-Dirac}) and the four-point coupling of $H_R$. The
interaction rate is approximately given by
\begin{eqnarray}
 \Gamma \sim \xi_R \frac{v^2}{(4 \pi)^3 \Lambda^2} T\ .
\label{eq:rate}
\end{eqnarray}
By requiring the interaction rate in Eq.~(\ref{eq:rate}) to be smaller
than the expansion rate of the universe at the time of BBN ($T \sim
1$~MeV), we obtain a lower bound on the cutoff scale to be $\Lambda
\gtrsim 4 \times 10^8$~GeV with assuming ${\cal O}(1)$ quartic coupling
constant of $H_R$.  The bound corresponds to the discrete symmetry
breaking scale to be $\langle H_R \rangle \gtrsim 400$~keV\@. Since the
bound from the domain wall formation is $\langle H_R \rangle \lesssim
400$~keV, a small coefficient for the four-point $H_R$ coupling allows
to have a viable parameter region around $\Lambda \sim 10^8$~GeV\@.

If we assume that $R$-sector particles are in thermal equilibrium 
with standard model particles above the temperature of the QCD phase 
transition, the number density of $R$-sector particles
are only suppressed by 1/10 compared to the photon number density. In
this case, a more severe constraint $\Lambda \gtrsim 10^{10}$~GeV is
obtained.

For $n = 2$, $R$-sector particles are most effectively produced 
through dimension six operators Eq.~(\ref{D6-eq}) and
\begin{eqnarray}
 \frac{1}{\Lambda^2} 
\bar{\nu}_L \gamma^{\mu} \nu_L H_R^* \partial_\mu H_R\ .
\end{eqnarray}
Again, a small amount of the $R$-sector particles is enough to quickly
thermalize through the processes $\nu_L H_R \to \bar{\nu}_R
H_R^{*}$ and $\nu_L \nu_R \to H_R^* H_R^*$ via the dimension five
operator
\begin{eqnarray}
 \frac{v}{\Lambda^2} \nu_L \nu_R H_R^2 \ .
\end{eqnarray}
Since the interaction rate of this process is larger at higher temperature, 
we require that these processes are already frozen out by the time
the Universe cools to the temperature of the QCD phase transition 
$T \sim 200$~MeV\@.  We obtain the bound $\Lambda \gtrsim 30$~TeV 
($\langle H_R \rangle \gtrsim 30$~MeV) and thus this model is excluded
by the domain wall constraint.

The models with $n \geq 3$ are obviously also excluded by domain wall
formation since the symmetry breaking scale is larger than 100~MeV for
$\Lambda \gtrsim 1$~TeV\@.

We conclude that BBN and domain wall formation severely constrains 
the model.  The only non-excluded region we found is around 
$\Lambda \sim 10^8$~GeV, leaving intact a large hierarchy between the
cutoff scale and the electroweak scale.

\subsection{A New $U(1)$ Gauge Symmetry}

Since we could not find interesting Dirac neutrino models with discrete
symmetries, we then consider the models with a continuous gauged symmetry
which are free from the domain wall problem.
We assume right-handed neutrinos $\nu_R^i$, $i=1, \cdots, n_R$ carry
$U(1)_R$ charges while quarks and leptons do not, because the observed
matter does not appear to have any gauge interactions beyond the usual
$SU(3)_C \times SU(2)_L \times U(1)_Y$.  The right-handed neutrino
charges $Q = (q_1, \cdots, q_{n_R})$ are subject to anomaly cancellation
conditions,
\begin{eqnarray}
  \label{eq:5}
  \sum_i q_i &=& 0, \label{grav-anom-eq} \\
  \sum_i q_i^3 &=& 0. \label{triple-anom-eq}
\end{eqnarray}
Also, since the right-handed neutrinos should carry $Z^\ell_3$
charge $-1$ to form Dirac masses, we need at least three $\nu_R$'s
to cancel the gravitational anomaly.

There are additional requirements on the $U(1)_R$ charge
assignments.  Ordinary dimension-4 Dirac neutrino masses $L
H \nu_R$ should be forbidden by gauge symmetry, since the purpose
of the gauge symmetry is to, at least ideally, avoid requiring
unnaturally small Yukawa couplings.  This means
at least three neutrinos must have
$U(1)_R$ charge.\footnote{If we allow to introduce another
discrete symmetry $Z_2$, $Q=(+1, -1, 0)$ is possible with
assigning $Z_2$ odd for $\nu_R^3$. In this case, one of the three
neutrinos remains massless.} Given just one scalar field
that breaks the $U(1)_R$ symmetry, the neutrino
$U(1)_R$ charges should be equal or opposite to that of the
breaking field.

One obvious choice of $U(1)_R$ charges for the neutrinos is
\begin{center}
\begin{tabular}{r|cccc}
           & $\nu^1_R$ & $\nu^2_R$ & $\nu^3_R$ & $\nu^4_R$ \\ \hline
$Q_R$      &   $+1$    &   $-1$    &   $+1$    &   $-1$   \\
$Z_3^\ell$ &   $-1$    &   $-1$    &   $-1$    &   $0$
\end{tabular}
\end{center}
Note that Majorana mass terms for vanishing $U(1)_R$ charge
combinations of $\nu_R$'s are forbidden by the unbroken $Z_3^\ell$
symmetry.  Also, for $Z_3^\ell$ to be anomaly free, three
neutrinos must have $Z_3^\ell$ charge $-1$ while the other is
neutral. The neutral one, $\nu^4_R$, can obtain a Majorana mass
after the $U(1)_R$ breaking.

There are other possible $U(1)_R$ charge assignments, including
chiral ones. In fact, as we show in Appendix \ref{sec:charges},
there are no integer solutions to both requirements,
Eqs.~(\ref{grav-anom-eq}) and (\ref{triple-anom-eq}), with up to
and including five right-handed neutrinos.  With six neutrinos,
one can find many solutions: $Q=(1, 1, 1, -4, -4, 5)$, $(1, 1, 5,
-9, -9, 11)$, etc. We expect the phenomenology of the model is
more or less the same for such cases.

The Dirac neutrino masses are generated by the U(1)$_R$ symmetry
breaking in low energy. If we assume U(1)$_R$ is broken by the vev of a
Higgs field $H_R$ with charge $-1/n$, we can write down an operator 
similar to that in Eq.~(\ref{eq:Dirac-op}) 
\begin{eqnarray}
 {\cal L}^R = \lambda^{\alpha i}
\left( \frac{H_R^{(*)}}{\Lambda} \right)^n
(H L^\alpha) \nu_R^i \ ,
\label{eq:Dirac-op2}
\end{eqnarray}
where complex conjugation for $H_R$ is necessary for the generations
with $Q_R = -1$.
The U(1)$_R$ breaking scale is the same as the discrete case and given
in Eq.~(\ref{eq:HRscale}).

In the case of the charge assignment $Q=(+1, -1, +1, -1)$, one of
the right-handed neutrinos with vanishing $Z_3^\ell$ charge
acquires a Majorana mass through the operator $(\nu_R^4)^2
H_R^{2n}/\Lambda^{2n-1}$. In the chiral case with
$Q=(1,1,1,-4,-4,5)$, one neutrino with charge $+5$ and one of the
two neutrinos with charge $-4$ couple to $H_R^n$ to become a Dirac
neutrino with mass of order $\vev{H_R}^n/\Lambda^{n-1}$.  The
other right-handed neutrino with charge $-4$ may have small
Majorana masses of the order of $\vev{H_R}^{8n}/\Lambda^{8n-1}$.

After $U(1)_R$ symmetry breaking, the gauge field $A^R_\mu$
acquires a mass $m_A \sim q_R g_R \vev{H_R}$, where $q_R$ is the
$U(1)_R$ charge of $H_R$.  From Eq.~(\ref{eq:HRscale}), we expect
$\langle H_R\rangle$ to be at least of order 1~eV\@.  Therefore $m_A
\gtrsim 10$~eV (assuming ${\cal O}(1)$ coupling constants), and
hence $A^R_\mu$ does not mediate a long range force.

\subsection{Kinetic Mixing}

The neutrino sector and the electroweak sector are not completely
decoupled.  Kinetic mixing between U(1)$_R$ and U(1)$_Y$ arises
at the renormalizable level, resulting in new interactions between
$A^R_\mu$ and charged fermions of the SM.  We parameterize the
effective size of the mixing by $\varepsilon$ that we will treat as a
perturbation, i.e., $\varepsilon \ll 1$.

Below the electroweak breaking scale, the gauge kinetic terms for
the unbroken gauge symmetries are
\begin{equation}
{\cal L}^A = -\frac{1}{4}(F_{\mu\nu}F^{\mu\nu} + F^R_{\mu\nu}
F^{R\mu\nu} - 2 \, \varepsilon F_{\mu\nu}F^{R\mu\nu}),
\label{LA1}
\end{equation}
where $F^{(R)}_{\mu\nu} \equiv \partial_\mu A^{(R)}_\nu -
\partial_\nu A^{(R)}_\mu$.  We consider the following redefinition
of the fields\footnote{We note that $U(1)_R$ will be spontaneously
broken, resulting in a massive vector boson.  This redefinition is
the only one that keeps the mass term for $A_\mu^R$ diagonal, and
hence is the canonical choice.}
\begin{equation}
A_\mu \to A_\mu + \varepsilon A^R_\mu \, \, ; \, \, A^R_\mu \to
A^R_\mu,
\label{Redef}
\end{equation}
resulting in
\begin{equation}
{\cal L}^A = -\frac{1}{4}(F_{\mu\nu}F^{\mu\nu} + F^R_{\mu\nu}
F^{R\mu\nu}) + {\cal O}(\varepsilon^2).
\label{LA2}
\end{equation}
With the gauge kinetic terms canonical, the charged fermions now
couple to $A^R_\mu$ via
\begin{equation}
Q_f \, {\bar \psi} {\not\!\! A} \psi \to
Q_f \, {\bar \psi} {\not\!\! A} \psi +
\varepsilon \, Q_f \, {\bar \psi} {\not\!\! A_R} \psi.
\label{Rcharge}
\end{equation}
Hence, $A_\mu$ is identified as the usual photon, however,
fermions of electric charge $Q_f$ have picked up a charge
$Q_R = \varepsilon Q_f$ under $U(1)_R$.

\begin{figure}[t]
\centerline{\includegraphics[width=0.9\hsize]{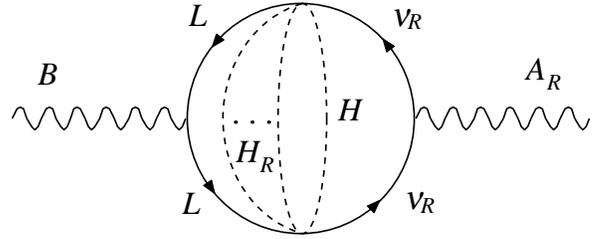}}
\caption{(2+$n$)-loop diagram that radiatively induces kinetic mixing
between $U(1)_Y$ and $U(1)_R$.} \label{nloop-fig}
\end{figure}

What is the size of the kinetic mixing $\varepsilon$?
Above the electroweak scale we can write
\begin{equation}
\left( \varepsilon_0 + \varepsilon_2 \frac{H^\dag H}{\Lambda^2}
+ \ldots \right) F_{\mu\nu}^Y {F^{\mu\nu}}^R \; ,
\label{eq:epsilon}
\end{equation}
where $\varepsilon_i$ correspond to the coefficients of the $i+4$
dimensional operators.  The effective value evaluated at the cutoff
scale, $\varepsilon(\Lambda)
\equiv \varepsilon_0 + \varepsilon_2 v^2/\Lambda^2 + \ldots$,
renormalizes to a low energy value $\varepsilon(\mu)$ through
radiative corrections that appear at (2+$n$)-loops, see
Fig.~\ref{nloop-fig}.
We estimate the loop-induced kinetic mixing between $B$ and $A_R$ to be
\begin{equation}
\varepsilon_{rad} \sim \frac{g_Y g_R}{(16 \pi^2)^{2+n}} \; .
\label{epsrad}
\end{equation}
where $g_Y = e/\cos\theta_W$ is the $U(1)_Y$ coupling and $g_R$ the
$U(1)_R$ coupling.  (At the electroweak scale this trivially matches to
the kinetic mixing between the photon and $A_R$,
$\varepsilon \sim (16 \pi^2)^{-2-n} e g_R$.)
Hence, it is technically natural to assume that the kinetic mixing at
the cutoff scale is of the order of $\varepsilon(\Lambda) \sim
(16 \pi^2)^{-2-n} g_Y g_R$.  An appropriate UV completion could justify
this assumption, for example by embedding $U(1)_R$ into an $SU(2)_R$
that is broken down to $U(1)_R$ at a low scale $\mu \ll \Lambda$, but
we will not pursue this further.  Instead, we will simply assume the
cutoff scale value is zero and that the leading order contribution to
the kinetic mixing arises only from the loop contributions. We now
proceed to examine phenomenological constraints on the parameters of
the model.

\subsection{Phenomenological Constraints\label{sec:pheno}}

We now consider several experimental bounds on the size of the
kinetic mixing $\varepsilon$.

\subsubsection{Precision data}

Precision data on $g-2$ of the electron agree with the SM to
better than 1 part in $10^{12}$.
For $m_A \ll m_e$, this suggests that the one-loop contribution with $A_R$
in place of a photon is $\varepsilon^2 \alpha/\pi$,
and therefore $\varepsilon \lesssim 10^{-4}$ which is satisfied for any $n$.
The case with $m_A \gg m_e$ is less constrained because of further
suppression of $m_e^2 / m_A^2$.

\subsubsection{Cosmic strings}

The spontaneous breaking of the U(1)$_R$ gauge symmetry leads to
formation of cosmic strings. The existence of the cosmic string causes
discontinuities in the temperature of the CMB\@. The bound of $\Delta T/T
\lesssim 10^{-5}$ puts a constraint on the breaking scale: $\vev{H_R}
\lesssim 10^{-2.5} M_P$. By comparing this with Eq.~(\ref{eq:HRscale}),
one finds it does not give a useful bound as long as we are interested
in low-cutoff theories.

\subsubsection{BBN}

The number of relativistic degrees of freedom during BBN is 
constrained by the effective number of neutrino species 
$\Delta N_\nu \lsim 1$ as we discussed with in the discrete 
Dirac neutrino case, Sec.~\ref{Dirac-discrete}.  All of the constraints 
that we found arising from the Yukawa interaction also apply to
the continuous U(1)$_R$ symmetry case.

Here we discuss additional constraints from the new gauge interaction.
Since the QCD phase transition is sufficient to dilute any pre-existing
abundance of $R$-sector particles, we need only consider processes
that are in equilibrium below the temperature of the QCD phase transition.
To delineate the restrictions on the parameter space, each production 
process can then be further divided into three cases depending on the
mass of the gauge boson.
 
\begin{figure}[t]
  \centering \includegraphics[width=\columnwidth]{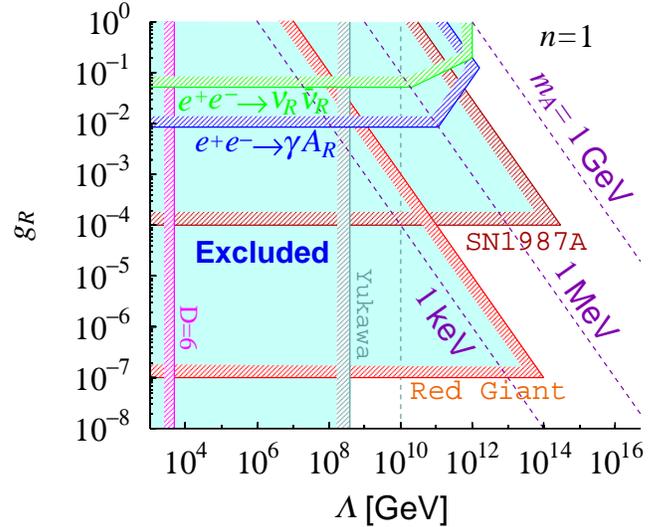}
  \caption{The astrophysical constraints for the case $n=1$ are shown in
  the $(\Lambda, g_R)$ plane.  The shaded region on the left-hand side
  is excluded.  The constraints come from $e^+ e^- \rightarrow \nu_R
  \bar{\nu}_R$ [Eqs.~(\ref{eq:nuRpair})]; $e^+ e^- \rightarrow \gamma
  A_R$ [Eqs.~(\ref{eq:AR})];
    the $D=6$ operator Eq.~(\ref{D6-eq});
    the Yukawa mediated processes $\nu_L \nu_R \rightarrow H_R H_R^*
    H_R^* \,{\rm or} \, \nu_L H_R \to \bar{\nu}_R H_R H_R^*$;
    overcooling red giants [Eq.~(\ref{redgiantbound})]; and overcooling
    SN1987A [Eq.~(\ref{SNbound})].  The diagonal dashed lines
    show contours of constant $m_A$.  In the case of a discrete
    symmetry, all constraints except for Yukawa and $D=6$ disappear.  If
    we assume the $R$-sector particles are in thermal equilibrium with
    the standard model particles before the QCD phase transition, the
    Yukawa constraint moves to the dotted line.  } \label{excluded1-fig}
\end{figure}

\begin{figure}[t]
  \centering \includegraphics[width=\columnwidth]{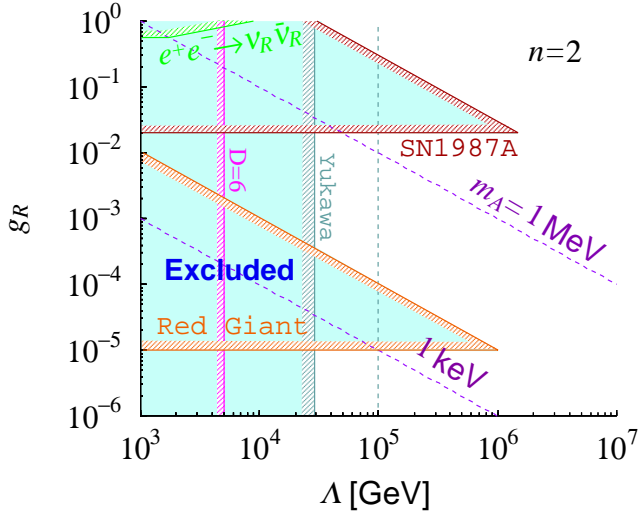}
  \caption{Same as Fig.~\ref{excluded1-fig} but for $n=2$ and 
    the scales are quite different. }
  \label{excluded2-fig}
\end{figure}

\begin{figure}[t]
  \centering \includegraphics[width=\columnwidth]{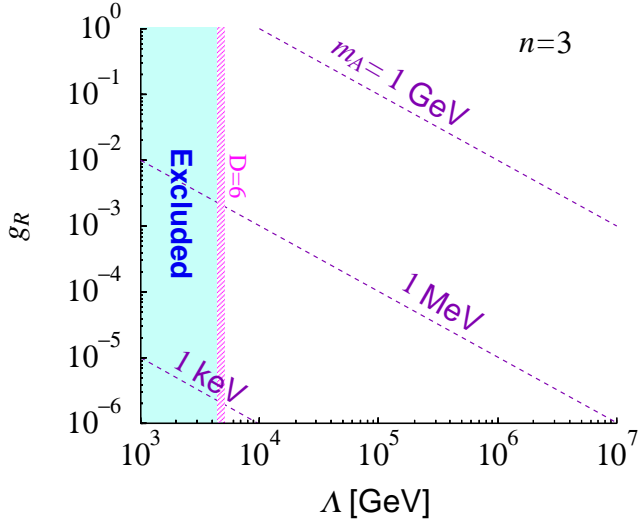}
  \caption{Same as Fig.~\ref{excluded2-fig} but for $n=3$.}
  \label{excluded3-fig}
\end{figure}

One important process is $e^+ e^- \to \nu_R \bar{\nu}_R$ through
$A_R$-exchange that occurs at a suppressed level proportional to the
amount of kinetic mixing.  We require that this process is not in
equilibrium at whatever temperature (between BBN to the QCD phase 
transition) yields the maximum interaction rate.  This gives the 
following constraints that depend on the gauge boson mass,
\begin{eqnarray}
g_R^2 \alpha \varepsilon^2 T \lesssim \frac{T^2}{M_P} 
  & \!\!\! & (m_A < 1 \; {\rm MeV}; T = 1 \; {\rm MeV}) \nonumber \\
\frac{g_R^2 \alpha \varepsilon^2 T^5}{m_A^4} \lesssim \frac{T^2}{M_P} 
  & \!\!\! & (1 \; {\rm MeV} < m_A < 200 \; {\rm MeV}; T = m_A) 
  \phantom{0000} 
  \label{eq:nuRpair} \\
\frac{g_R^2 \alpha \varepsilon^2 T^5}{m_A^4} \lesssim \frac{T^2}{M_P} 
  & \!\!\! & (m_A > 200 \; {\rm MeV}; T = 200 \; {\rm MeV}) \nonumber 
\end{eqnarray}
The resulting constraints are shown in 
Figs.~\ref{excluded1-fig}-\ref{excluded3-fig}.

The second process is the production of $A_R$ through for example 
$e^+ e^- \rightarrow \gamma A_R$.  Again, requiring that this process is not 
in equilibrium at the temperature that yields the maximum interaction rate
yields the following constraints that depend on the gauge boson mass,
\begin{eqnarray}
4\pi \alpha^2 \varepsilon^2 T \lesssim T^2/M_P & & 
  (m_A < 1 \; {\rm MeV}; T = 1 \; {\rm MeV}) \nonumber \\
4\pi \alpha^2 \varepsilon^2 T \lesssim T^2/M_P & & 
  (1 < m_A < 200 \; {\rm MeV}; T = m_A) 
  \phantom{00000} 
  \label{eq:AR} \\
\mbox{(no constraint)} & & 
  (m_A > 200 \; {\rm MeV}) \nonumber
\end{eqnarray}
The resulting constraints are again shown in
Figs.~\ref{excluded1-fig}-\ref{excluded3-fig}.

\subsubsection{Red Giant and Supernova Cooling}

Astrophysical bounds from excessive energy loss through the
emission of light, extremely weakly interacting particles from
stars and supernovae are potentially much more severe if
kinematically allowed.  This is well known to put significant
constraints on axions, and here we will use these bounds to
constrain the emission rate of $A_R$.

Consider the emission of a light axion $a$, coupled to a
fermion $\psi$ with strength $g_\psi$
\begin{equation}
{\cal L}^a = -i g_\psi \, {\bar \psi} \gamma_5 \psi \, a.
\label{La}
\end{equation}
Here, $g_\psi = m_\psi/f_a$, where $m_\psi$ is the fermion mass
and $f_a$ is the effective Peccei-Quinn scale.  Using
astrophysical data on red giants, one finds, for $\psi = e$, $g_e
\lesssim 2.5 \times 10^{-13}$, corresponding to roughly $f_a \gtrsim
10^9$~GeV \cite{Raffelt:1994ry,Raffelt:1999tx}.

Axion emission proceeds through the pseudoscalar interaction,
Eq.~(\ref{La}), that vanishes in the non-relativistic limit.
Emission of $A_R$, however, proceeds through the usual vector interaction
without any non-relativistic suppression.  This means that the amplitude
for the emission of $A^R_\mu$ off an electron is expected to be enhanced
by an amount $\sim m_e/T_{\rm giant} \sim 50$ with respect to axion emission,
where $T_{\rm giant} \sim 10$~keV is the core temperature of the red giant.
Using $\varepsilon \sim 50 g_e$, we obtain the stringent upper limit
$\varepsilon \lesssim 10^{-14}$ when $m_A \lsim T_{\rm giant}$.
It is straightforward to translate the constraints on the parameters
$\varepsilon$ and $m_A$ into constraints on $g_R$ and $\Lambda$.
We obtain the following exclusion regions:
\begin{equation}
\begin{array}{rccl}
n = 1: \quad & g_R \gsim 10^{-7} & \;\; {\rm or} \;\; &
   g_R \Lambda \lsim 10^7 \; {\rm GeV} \\
n = 2: \quad & g_R \gsim 10^{-5} & \;\; {\rm or} \;\; &
   g_R \Lambda \lsim 10 \; {\rm GeV}
\end{array}
\label{redgiantbound}
\end{equation}
for the two cases.  The excluded region for the cases $n=1,2$ are shown
as the lower triangular regions in
Figs.~\ref{excluded1-fig},\ref{excluded2-fig}.  Given the range of 
scales shown in Fig.~\ref{excluded3-fig}, there is no constraint
for $n \geq 3$.

Are $A_R$'s ever strongly enough coupled to matter to be
trapped in red giants?  Assuming the scattering of $A_R$ is dominated
by Compton scattering, the cross section can be estimated as
$\sigma \sim \varepsilon^2 \alpha^2 m_e^{-2} \sim
\varepsilon^2 10^{-25}$~cm$^2$.  Using $n_e \sim 10^{30}$~cm$^{-3}$
as the average number density of electrons in the core, we obtain
$d \sim (n_e \sigma)^{-1} \sim \varepsilon^{-2} 10^{-5}$~cm.
This can be written as
\begin{equation}
d \sim \frac{10^{5 + 4 n} \; {\rm cm}}{g_R^2}
\end{equation}
This distance is larger than the core radius of a red giant,
$R_{core} \sim 10^{9}$~cm, regardless the dimension of the
operator or other parameters so long as $g_R \lsim 1$.

We note that assuming $g_R \gsim 1$ is not a viable solution either.  In
this case, we have $m_A \sim g_R \langle H_R \rangle \gsim
100$~MeV, given our previous bounds. However, in the examples that
we have considered, there is always a $\nu_R$ with small ($\ll
1$~eV) Majorana mass in the spectrum. The decay rate of $A_R$ is
then bounded from below by $g_R^2 m_A \gsim 100 {\rm MeV} \sim
10^{22}$~s$^{-1}$.  Thus, $A_R$ quickly decays into $\nu_R {\bar \nu_R}$
which, in turn, escape the star.  Hence, $A_R$ particles cannot be
trapped in a red giant in any regime of parameters.

The observation of neutrinos from SN1987A can also be used to
place bounds on the emission of axions and $A_R$. Axions are
emitted via bremsstrahlung off of nucleons in the nucleon
scattering process $N N \to N N a$ via pion exchange
\cite{Iwamoto:1984ir}. As a result, the axion-nucleon coupling
range $3 \times 10^{-10} < g_N < 3 \times 10^{-7}$ is ruled out
\cite{Eidelman:2004wy}. Values larger than the upper limit are allowed
for axions due to the possibility of trapping inside the
supernova.

Translating this bound on axions into a
bound on $\varepsilon$, there is again a relative enhancement of
the $A_R$ coupling to nucleons since it does not have a
non-relativistic suppression compared with axions.  The coupling
enhancement is roughly $m_N/{T_{\rm SN1987A}} \sim 30$, where
$T_{\rm SN1987A} \sim 30$ MeV, suggesting that the range $10^{-11}
< \varepsilon < 10^{-8}$ is excluded so long as $m_{A_R} \lsim
T_{\rm SN1987A}$.  This implies $g_R \gsim 0.1$ for $n \geq 1$
may be allowed for trapped $A_R$'s.  However, this is not the case.
In a similar fashion as in the case of red giants, we get
an upper bound on the lifetime of $A_R$ of order $10^{-20}$~s.
This is much shorter than the typical cooling timescales for
a supernova, which is of order seconds.  The fast decay $A_R \to
\nu_R {\bar \nu_R}$ then results in overcooling of the supernova
and trapping is again irrelevant here.
We thus obtain the following exclusion bounds:
\begin{equation}
\begin{array}{rccl}
n=1: \quad & g_R > 10^{-4} & \;\; {\rm unless} \;\;
  & g_R \Lambda > 10^{11} \; {\rm GeV} \\
n=2: \quad & g_R > 0.02 & \;\; {\rm unless} \;\;
  & g_R \Lambda > 50 \; {\rm TeV} \; .
\end{array}
\label{SNbound}
\end{equation}
The excluded region of parameters for $n=1, 2$ is shown in
Figs.~\ref{excluded1-fig} and ~\ref{excluded2-fig}, respectively.
There is no constraint for $n \geq 3$.

\section{Hierarchy\label{sec:hierarchy}}

Our models contain a (smaller) hierarchy problem of their own: the
hierarchy between the gauge symmetry breaking scales and the cutoff
scale.  In Majorana case, we considered symmetry breaking in the
infrared, in the range $2$--$400$~keV\@.  It is possible that this
scale could be stabilized by supersymmetry.  In the effective theory
below the electroweak scale, the scalar fields $\chi$ acquire a 1-loop
quadratically divergent contribution proportional to $g^2$.  If this
were the only contribution to the scalar (mass)$^2$, the hierarchy
would be stabilized.  However, there are cutoff suppressed
contributions such as $Z^\dag Z \chi^\dag \chi/\Lambda^2$, where $Z$
is a hidden sector field with an $F$-term, that are not forbidden by
symmetries.  One possibility is to remove these operators through
sequestering, thereby allowing anomaly mediation to dominate
\cite{AMSB}.  One might be concerned about the quadratic divergence of
the scalar masses below the supersymmetry breaking scale.  This arises
at two- or three-loops, giving a contribution at most of order
$m_{\chi}^2 \sim m_{\rm SUSY}^6/((16 \pi^2)^3 \Lambda^4)$.  But, it is
easy to see that so long as the scale of the soft breaking $m_{\rm
  SUSY} \lsim 4 \pi v$, this higher order contribution is subdominant
to the already safe one-loop quadratically divergent contribution
below the electroweak breaking scale discussed above.

Another way to stabilize this hierarchy is to use technicolor.
Consider the Dirac case for purposes of illustration.
Suppose there is a new $SU(3)$ gauge group
with dynamical scale $\Lambda_R$, that couples to two triplets $q_i
({\bf 3}_0)$ with $i=1,2$ and two anti-triplets $\bar{q}_+
(\overline{\bf 3}_{+1})$ and $\bar{q}_- (\overline{\bf 3}_{-1})$.  The
subscripts here refer to the $U(1)_R$ charges. Note that the gauge
invariance completely forbids mass terms for techniquarks.  The theory
is identical to the two-flavor QCD and is known to cause a condensate
$\langle \bar{q}_+ q_1 \rangle = \langle \bar{q}_- q_2 \rangle \simeq
\Lambda_R^3 \neq 0$ that breaks $U(1)_R$.
This model corresponds to $n=3$ case in Eq.~(\ref{eq:Dirac-op2}) with
$H_R^3$ replaced by the fermion bilinear
\begin{equation}
{\cal L}^R = \lambda^{\alpha i} \frac{(H L^\alpha)[(\bar{q}_- q_2)
\nu^i_R]}{\Lambda^3}\ . \label{eq:technicolor}
\end{equation}
While this mechanism stabilizes $\Lambda_R$ against $\Lambda$,
and naturally leads to the small Dirac neutrino masses, the
U(1) symmetry breaking scale is much larger,
$\Lambda_R \sim (m_\nu/v)^{1/3} \Lambda \gsim 10^{-4} \Lambda$.
Choosing a cutoff $\Lambda \gsim 100$ TeV (to avoid FCNC problems
in the SM) implies $\Lambda_R \gsim 10$ GeV\@.

We note that lowering the cutoff scale of the Standard Model may also
be constrained by too-large flavor changing neutral current processes
from cutoff scale suppressed flavor-violating operators.  The Majorana
case and the $n \le 2$ Dirac cases are roughly safe from these
constraints, since we found the cutoff scale must be larger than tens
of TeV\@.  (This is curiously reminiscent of the smallest cutoff scale
that was found by using split fermions as a solution to global
symmetry breaking \cite{Arkani-Hamed:1999dc}.)  For $n \ge 3$
in the Dirac case, the flavor problem becomes more acute as the cutoff
scale is taken to the lowest possible value, of order 5 TeV\@.

\section{Conclusion\label{sec:conclusion}}

The hierarchy problem may be solved by lowering the cutoff scale
of the SM.  However, this leads to unacceptably large violations
of baryon and lepton numbers.  We have shown there are simple
discrete gauge symmetries that can be imposed on the SM to protect
against rapid proton decay and too-large Majorana neutrino mass.
We then constructed models that generate small neutrino mass
through spontaneous (discrete or continuous) gauge symmetry
breaking at small scales.

In our first model, Majorana neutrino mass were generated once
$Z_3^\ell$ is broken.  This model is restricted by the experimental
constraints from the
non-observation of $0\nu\beta\beta\chi$, which requires $\Lambda >
30$ TeV, and the non-observation of modifications to $\Delta T/T$
in the CMB from a (non-frustrated) domain wall network, which
requires $\Lambda \lesssim 400$ TeV\@. Generating neutrino masses
near the observed size implies the scale for $Z_3^\ell$ breaking
is of order $2$--$400$ keV\@.  Values near the upper end of this range
yield domain walls in the universe that modify the CMB at
observable levels.  If the domain wall network were frustrated,
such as if $Z_3^\ell$ came from a larger discrete symmetry, the
domain walls could provide the dark energy of the universe with
$w = -2/3$.

Dirac-neutrino masses arise when $Z_3^\ell$ is exact and right-handed
neutrinos are added to the model.  We proposed a new discrete and
$U(1)_R$ symmetry that acts only on right-handed neutrinos.  Dirac
neutrino masses of the observed size are generated once the symmetry is
broken.  Limiting the light degrees of freedom during BBN gives the 
strongest constraint on the models.
We found that the models with a discrete symmetry are disfavored by the
domain wall constraint.  Models with a U(1), however, are viable with 
$\Lambda \gtrsim 5$-$30$ TeV when the dimension of the operator that
generates the Dirac neutrino masses is more than six ($n \geq 2$).
More precise constraints are shown in the figures.  The new gauge boson
mass is $m_A \simeq g_R \mbox{100 MeV}$ for $n=2$ and higher for larger
$n$.  If the gauge coupling is small, a very light gauge boson is
possible.

\appendix

\section{Alternative Discrete Lepton Number}
\label{alternative-appendix}

Note that flavor-blindness in the lepton sector can be relaxed with
interesting consequences for allowed FCNC operators.  For example, the
assignment
\begin{center}
\begin{tabular}{l|cccccc}
           & $L_e$ & $e_e^c$ & $L_\mu$ & $e_\mu^c$ & $L_\tau$ & $e_\tau^c$
\\ \hline
$Z_5^\ell$ &  $-2$ &   $2$   &    $1$  &    $-1$   &    $1$   &   $-1$
\end{tabular}
\end{center}
forbids not only Majorana masses for neutrinos, but also the
flavor-changing lepton decay $\mu \to e \gamma$.  Majorana
neutrino masses with arbitrary flavor mixing can be generated with
two fields $\chi_{1,2}$ with charges $-1,-2$ that acquire vevs and
spontaneously break the $Z_5^\ell$ symmetry.  The process $\mu \to
e\gamma$ also reappears, but now further suppressed by $\langle
\chi_i \rangle/\Lambda$.
While we find this an amusing option, we adopt the flavor-blind
$Z_3^\ell$ in the body of the paper.

\section{$U(1)$ Charges \label{sec:charges}}

In this appendix, we show that there are no anomaly-free charge
assignments with fewer than six neutrino species that are (1) chiral,
and (2) two of the charges are the same.

It is easy to show that there are no chiral charge assignments up to
four neutrinos.  Without a loss of generality, we can always choose
one of the charges to be $+1$.  With four neutrinos, the charges are
$Q=(1, \alpha, \beta, \gamma)$.  The anomaly freedom requires
\begin{eqnarray}
  \label{eq:anomaly}
  1 + \alpha + \beta + \gamma &=& 0, \\
  1 + \alpha^3 + \beta^3 + \gamma^3 &=& 0.
\end{eqnarray}
Solving for $\gamma$ in the first equation and substituting into the
second, we find
\begin{equation}
  \label{eq:2}
  -3(1+\alpha)(1+\beta)(\alpha+\beta)=0.
\end{equation}
All solutions to this equation give vector-like charge assignments:
For $\alpha=-1$, we find $Q=(1, -1, \beta, -\beta)$.  For $\beta=-1$,
we find $Q=(1, \alpha, -1, -\alpha)$.  For $\alpha=-\beta$, we find
$Q=(1, \alpha, -\alpha, -1)$.

With five neutrino species, one can find non-trivial chiral charge
assignments.  There are integer solutions, such as $Q=(1, 5, -7, -8,
9)$ or $(2, 4, -7, -9, 10)$.  However, with only one Higgs that breaks
the $U(1)_R$ gauge symmetry, we need at least two neutrino species to
have the same charge so that both of them can acquire masses.  Without
a loss of generality, we can normalize the charges such that two
species have charge unity, $Q = (1, 1, \alpha, \beta, \gamma)$.  Then
the anomaly freedom requires
\begin{eqnarray}
  \label{eq:anomaly2}
  2 + \alpha + \beta + \gamma &=& 0, \\
  2 + \alpha^3 + \beta^3 + \gamma^3 &=& 0.
\end{eqnarray}
Substituting the solution for $\alpha$ from the first equation to the
second, we find
\begin{eqnarray}
  \label{eq:3}
  \alpha &=& - \frac{\gamma^2+4\gamma+4 \pm
    \sqrt{\gamma(\gamma^3-8\gamma-8)}}{4+2\gamma}, \\
  \beta &=& - \frac{\gamma^2+4\gamma+4 \mp
    \sqrt{\gamma(\gamma^3-8\gamma-8)}}{4+2\gamma}.
\end{eqnarray}
Therefore, for any choice of $\gamma$, we can always find appropriate
$\alpha$ and $\beta$.

However, there are no integer solutions.  The integer solutions would
require that $\alpha$, $\beta$, and $\gamma$ are all rational numbers.
We can always write $\gamma = p/q$, where $p$ and $q$ are relatively
prime.  In order for $\alpha$ and $\beta$ to be also rational, the
argument of the square root
\begin{equation}
  \label{eq:4}
  \gamma(\gamma^3-8\gamma-8) = \frac{p(p+2q)(p^2-2pq-4q^2)}{q^4},
\end{equation}
must be a complete square.  Therefore, the numerator on the r.h.s.
must be a complete square of an integer.  Because there is a
factor of $p$ already, other two factors must contain a factor of
$p$.  However, both $p+2q$ and $p^2-2pq-4q^2$ are relatively prime
with $p$ because $q$ is relatively prime with $p$.  Q.E.D.

\acknowledgments

H.D. was supported in part by the P.A.M. Dirac Fellowship,
awarded by the Department of Physics at the University of Wisconsin-Madison.
This work was supported by the IAS, funds for Natural Sciences,
as well as in part by the DOE under contracts
DE-FG02-95ER40896,
DE-FG02-90ER40542 and
DE-AC03-76SF00098, and in part by NSF grants PHY-0098840 and PHY-0070928.


\end{document}